\documentclass[%
 reprint,	
showpacs,
preprintnumbers,
amsmath,
amssymb,
aps,
prb,
floatfix,
]{revtex4-1}

\usepackage[english]{babel}
\usepackage{pgfplots}
\usepackage{placeins} 
\usepackage[caption=false]{subfig}
\usepackage{comment}
\usepackage{graphicx}
\usepackage{dcolumn}
\usepackage{bm}

\newcommand{\figref}[1]{Fig.~\ref{#1}}

\newcommand{\eg}{{\em e.\,g.}}
\newcommand{\ie}{{\em i.\,e.}}

\begin{document}


\title{Simple and efficient way of speeding up \\transmission calculations with $k$-point sampling}
\author{Jesper Toft \surname{Falkenberg}}
\affiliation{Center for Nanostructured Graphene (CNG), Department of Micro- and Nanotechnology (DTU Nanotech), Technical University of Denmark, DK-2800 Kongens Lyngby, Denmark}
\author{Mads \surname{Brandbyge}}
\email{mads.brandbyge@nanotech.dtu.dk}
\affiliation{Center for Nanostructured Graphene (CNG), Department of Micro- and Nanotechnology (DTU Nanotech), Technical University of Denmark, DK-2800 Kongens Lyngby, Denmark}
\date{\today}

\begin{abstract}
The transmissions as functions of energy are central for electron or phonon transport in the Landauer transport picture. 
We suggest a simple and computationally ``cheap'' post-processing scheme to 
interpolate transmission functions over $k$-points to get smooth well-converged average transmission functions. 
This is relevant for data obtained using typical ``expensive'' first principles calculations where the leads/electrodes are described by periodic boundary conditions. 
We show examples of transport in graphene structures where a speed-up of an order of magnitude is easily obtained. 
\end{abstract}

\maketitle

\section{Introduction}
Calculations of electronic conductance based on first principle methods such as density functional theory (DFT) provide a valuable tool in order to gain insights into electronic transport in nano-conductors and comparison to experiments without employing fitting parameters. This is for example the case in the field of single-molecular devices\cite{cuevas_2010_molecular}. Popular methods are based on DFT in combination with the non-equilibrium Green's function approach (DFT-NEGF), see e.g. \onlinecite{brandbyge_density_2002,rocha_2005_towards,strange_2007_benchmark}, or scattering wave-function approaches\cite{garcia-lekue_2010_plane}. The electrodes in such calculations are typically treated employing periodic boundary conditions in the direction transverse to the transport direction with a corresponding $k$-point average of the electronic states and transmissions. This means that for each transverse $k$-point the system essentially behaves as a one-dimensional conductor with diverging density of states and discontinuities in the transmission function at energies corresponding to band on-sets/channel openings. It is well known that often in order to obtain smooth, well-converged density of states and transmissions as a function of energy, a substantial number of transverse $k$-points are needed due to the rapid variations of these functions for individual $k$-points. 
Certain quantities, for example the Seebeck coefficient and thermo-electric figure of merit (ZT), are based on the detailed behavior of the transmission\cite{paulsson_thermoelectric_2003,zotti_2014_heat} and thus exceedingly sensitive to energy and $k$-resolution of the calculations. This can amount to a significant computational burden for large systems treated by first principle methods. Thus it is highly interesting to devise simple ways to make this more efficient and get maximum information from the data.

Sophisticated methods to tackle this include the transformation to a smaller basis-set using maximally localized Wannier functions\cite{marzari_maximally_1997}, or to construct a optimized minimal basis-set and using this to determine the transmission \cite{pizzi_boltzwann_2014}. Both require one to examine the details of the chemical bonds in the system, relevant energy windows, and storing wavefunctions, which tend to be elaborate. In this paper we present a simple and efficient post-processing interpolation scheme which can significantly speed up the convergence with respect to $k$-points. We illustrate the method by applying it to various graphene-based nano-structures which are prone to bad convergence due to its vanishing density of states at the Fermi level. 

In the remaining parts of the paper we first explain the workings of the interpolation scheme in Sec.~\ref{sec:description}, while we investigate various test cases in Sec.~\ref{sec:examples}, and finally discuss limitations to the scheme and conclude in Sec.~\ref{sec:conclusion}.

\begin{figure}[b]
	\centering
	\includegraphics{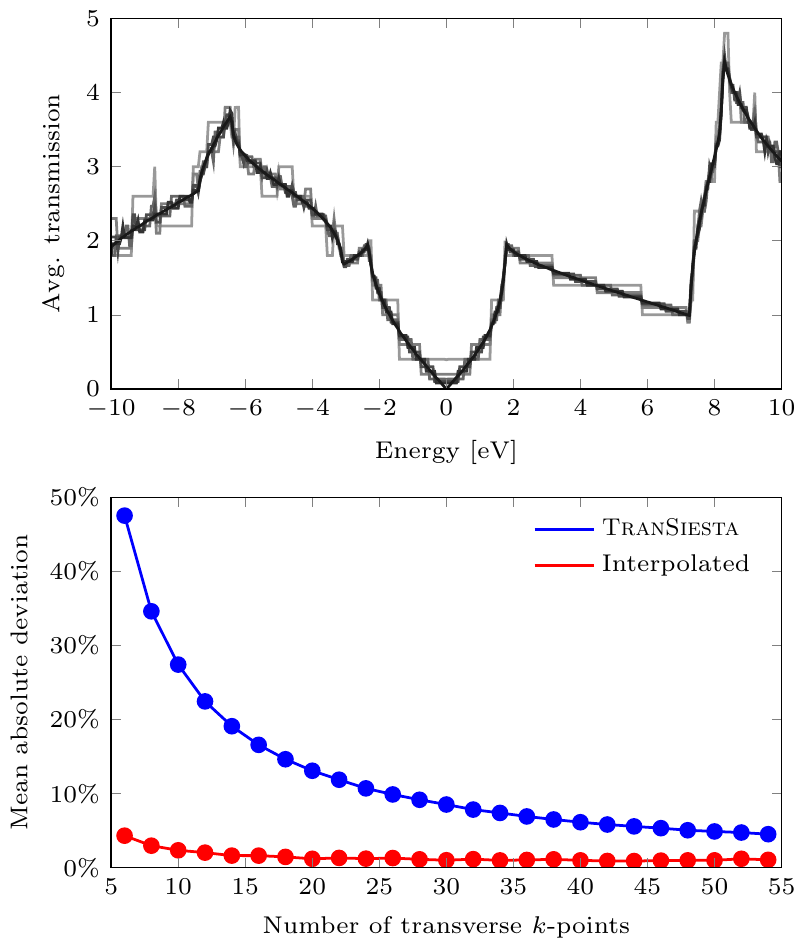}
	\caption{Analysis of transmission in pristine graphene: (a) Transmission converges as the number of transverse $k$-points increases (darker colors). (b) The mean absolute deviation from fully converged data shown for \textsc{TranSiesta} data before and after using our interpolation scheme.}
	\label{fig:introduction}
\end{figure}

\begin{figure*}[tb]
	\subfloat[Data transformation]{
		\centering
		\includegraphics{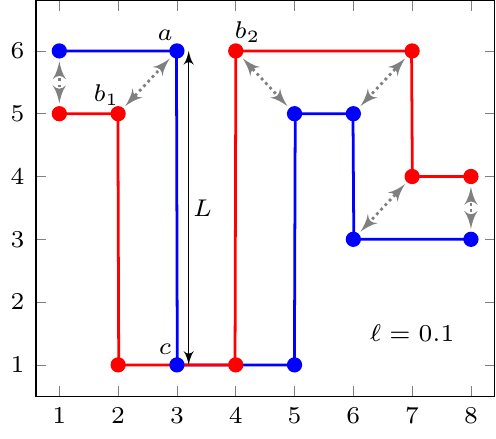}
	}
	\subfloat[Matrix construction]{
		\centering
		\includegraphics{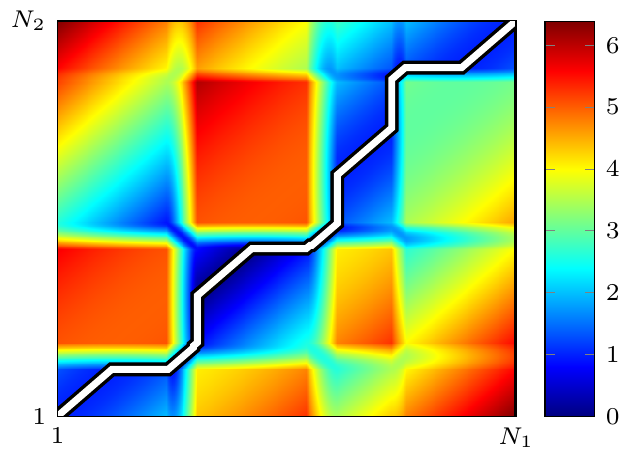}
	}
	\subfloat[Interpolation]{
		\centering
		\includegraphics{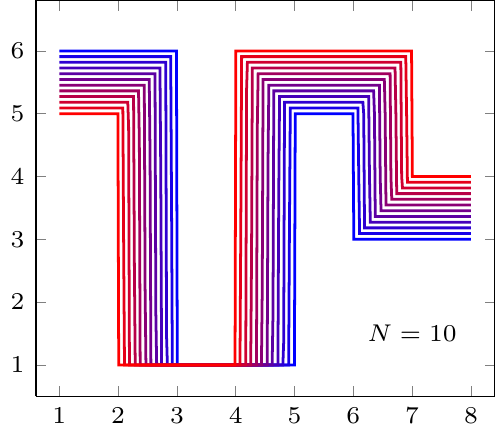}
	}
	\caption{(Colour online) We seek an interpolation of the two datasets (Red, Blue) containing $N_1$ and $N_2$ points, respectively. The three steps in the algorithm: (a) The original data sets are transformed into line segments of a maximum length $\ell=0.1$. Thus, for example the line segment $a$-$c$ is split into 50 points. (b) The weighted Euclidean distance is calculated for all point on both curves, and the shortest path from (1,1) to ($N_1,N_2$) is found. (c) The curves are linearly interpolated using the found path with $N=10$ intermediate curves.}
	\label{fig:allsteps}
\end{figure*}

\section{Description of the method}
\label{sec:description}

The use of computationally ``expensive'' first principles DFT-NEGF calculations for determining the transmission through nano-structured systems is limited by the amount of time one can afford to spend on the $k$-grid resolution. Often the rough behavior of the transmission is already seen with a limited number of $k$-points but the convergence of the average is slow since the functions are changing abruptly with energy, \eg\ around a band onset or a resonance. The position of the abrupt feature will typically shift in a smooth way with changing $k$-point, but a linear interpolation of the curves between two consecutive $k$-points will be of little use since it will simply contain, say, half of each abrupt feature.
Instead, we propose an interpolation scheme which can make use of a coarse, non-converged $k$-grid, and thus reduce the computation cost simply by using a ``clever'' technique to approximate the transmission curves for intermediate $k$-points. The method does not magically guess the correct interpolated curves, and one has to have a reasonable amount of $k$-point resolved transmission curves in order to obtain a useful result, but smooth averaged curves can be obtained, as we will illustrate below, using a significantly smaller number of $k$-points. The interpolation is done by using a shortest-path solver to determine a correspondence between two $k$-adjacent curves. The correspondence is then used to find intermediate curves which can be used to determine the averaged transmission. The proposed interpolation scheme consists of three separate steps, which will be described in detail in the following sections.

In order to show the validity of our proposed scheme we have determined the transmission through pristine graphene for increasing number of transverse $k$-points (see \figref{fig:introduction}a). The computational details are given in Sec.~\ref{sec:examples}. We compare ${N\!=\!6,8,...,54}$ to a well-converged calculation with ${N\!=\!600}$ $k$-points. By applying our algorithm we significantly reduce the mean absolute deviation from a fully converged transmission, as shown in \figref{fig:introduction}b. We see that the interpolated transmissions converge much quicker than the raw data. Fitting data with a power law, we can estimate the amount of $k$-points needed in order to obtain a deviation of less than one percent, thus giving a speed-up factor of $\eta=220/41=5.37$.

\subsection{Transform of data}

In the following we will outline the method in general terms and denote the data points by $(x,y)$, corresponding to the transmission function data point, $(E,T_k(E))$, for given $k$-point and energy in the concrete examples.  
Initially, we transform the set of data points $(x,y)$ into points with a maximum Euclidian distance $\ell$. The new data points span line segments (when connected) on the curve $\mathcal L$ consisting of $N$ different individual points. In practice the transformation of data is done by looping through all segment lengths $L$ and inserting additional data points if $L>\ell$. 
Figure~\ref{fig:allsteps}a shows data where the distance between points $a,c$ is much larger than $\ell$, and thus requires extra points between points $a,c$. The optimal value of the segment length $\ell$ depends on the given data. A value close to the median of the original point distance is a good starting guess, and is chosen as the default value.

The two curves in \figref{fig:allsteps}a are required to smoothly interpolate into each other -- a process which can be estimated by hand (gray arrows).
The proper path is found by overall minimization of the interpolation distance, which will described in the following subsection.

\subsection{Correspondence between curves}
Given two data sets, \ie\ the curves $\mathcal L_1, \mathcal L_2$, we construct a matrix $\mathcal D$ containing the weighted Euclidean distances between points on opposite curves,
\begin{align}
	\mathcal D_{ij} = \sqrt{w_x\left(\tilde{x}_1^i-\tilde{x}_2^j\right)^2 + w_y\left(\tilde{y}_1^i-\tilde{y}_2^j\right)^2},
\end{align}
where $i=1 \ldots N_1, j=1 \ldots N_2$ denote all points on the curves $\mathcal L_1,\mathcal L_2$, respectively. Thus, the dimension of the matrix $\mathcal D$ is $N_1 \times N_2$. The weights $w_{x,y}$ are added to ensure a degree of tunability when interpolating. This is due to the fact that $x$ and $y$ are different quantities and thus have different scales.
In order to interpolate one curve into the other we need to determine the shortest path for all points on either curve. This is done by considering the distance matrix as a landscape and moving from the start $(1,1)$ to the destination $(N_1,N_2)$ using the smallest cost possible. This is essentially identical to finding the minimum energy path on an potential energy surface, which can be done using either string methods or the nudged elastic band approach\cite{henkelman_optimization_2008}.
We will instead use Dijkstra's algorithm, which loops through coordinates in the distance landscape and iteratively compares all found paths until the target point is found. Each iteration is done while keeping track of the cost and path. For more information see Ref.~\onlinecite{dijkstra_1959_anote}. The coordinates along the returned shortest path is saved in a correspondence matrix $\mathcal C$ containing two columns $(\mathcal C_1,\mathcal C_2)$, which is used in the final step of our routine. The distance matrix $\mathcal D$ in \figref{fig:allsteps}b shows the shortest path as a white line propagating along the distance landscape minimum (black/dark blue).

\subsection{Interpolation of curves}
Once the data is transformed and a correspondence matrix is obtained we interpolate the curves $\mathcal L_1,\mathcal L_2$. Instead of a simple linear interpolation we use the correspondence matrix to obtain intermediate curves. A simple linear, discretized interpolation is used, i.e.,
\begin{align}
	x_{12\alpha} &= \alpha x_1 \left[\mathcal C_1 \right] + (1-\alpha) x_2 \left[\mathcal C_2 \right]\\
	y_{12\alpha} &= \alpha y_1 \left[\mathcal C_1 \right] + (1-\alpha) y_2 \left[\mathcal C_2 \right],
\end{align}
where $\alpha$ is the interpolation parameter between zero (curve $\mathcal L_2$) and one (curve $\mathcal L_1$), and $x_i[M]$ means that we use the indices $M$ of the vector $x_i$. The found intermediate data sets $(x_{12\alpha},y_{12\alpha})$ can then be sampled back to the original grid.
In practice we loop over $\alpha$ to generate the different interpolated curves. Figure~\ref{fig:allsteps}c shows the data in \figref{fig:allsteps}a interpolated into $N=10$ intermediate curves and transformed back onto the original grid.


We note that the choice of weights $w_x,w_y$ depends on the input data scales. Changing the values can highly affect the outcome of the shortest-path solver, since the distance landscape is changed. Usually, it is advisable to rescale the data (using the weights) so that the two data ranges are comparable. Similarly, the length $\ell$ has to be chosen wisely: A large value can result in crude interpolations while a too small value makes the algorithm too time-consuming.

Finally, we note that the algorithm described in the previous subsections allows us to interpolate data in general. We can apply the algorithm specifically to transmissions and DOS by providing it with the needed data. In the case of \textsc{TranSiesta} and the utility \textsc{TBTrans} we need to weight the interpolated curves using $k$-grid weights and sum to obtain average transmission curves. In the case of 3D transport we have a 2D $k$-grid, which can be investigated using bilinear interpolation. In the following section we apply the interpolation scheme to example cases.

\section{Example cases}
\label{sec:examples}

\begin{figure}[t]
	\centering
	\subfloat[Graphene sheet]{\includegraphics[width=0.35\textwidth]{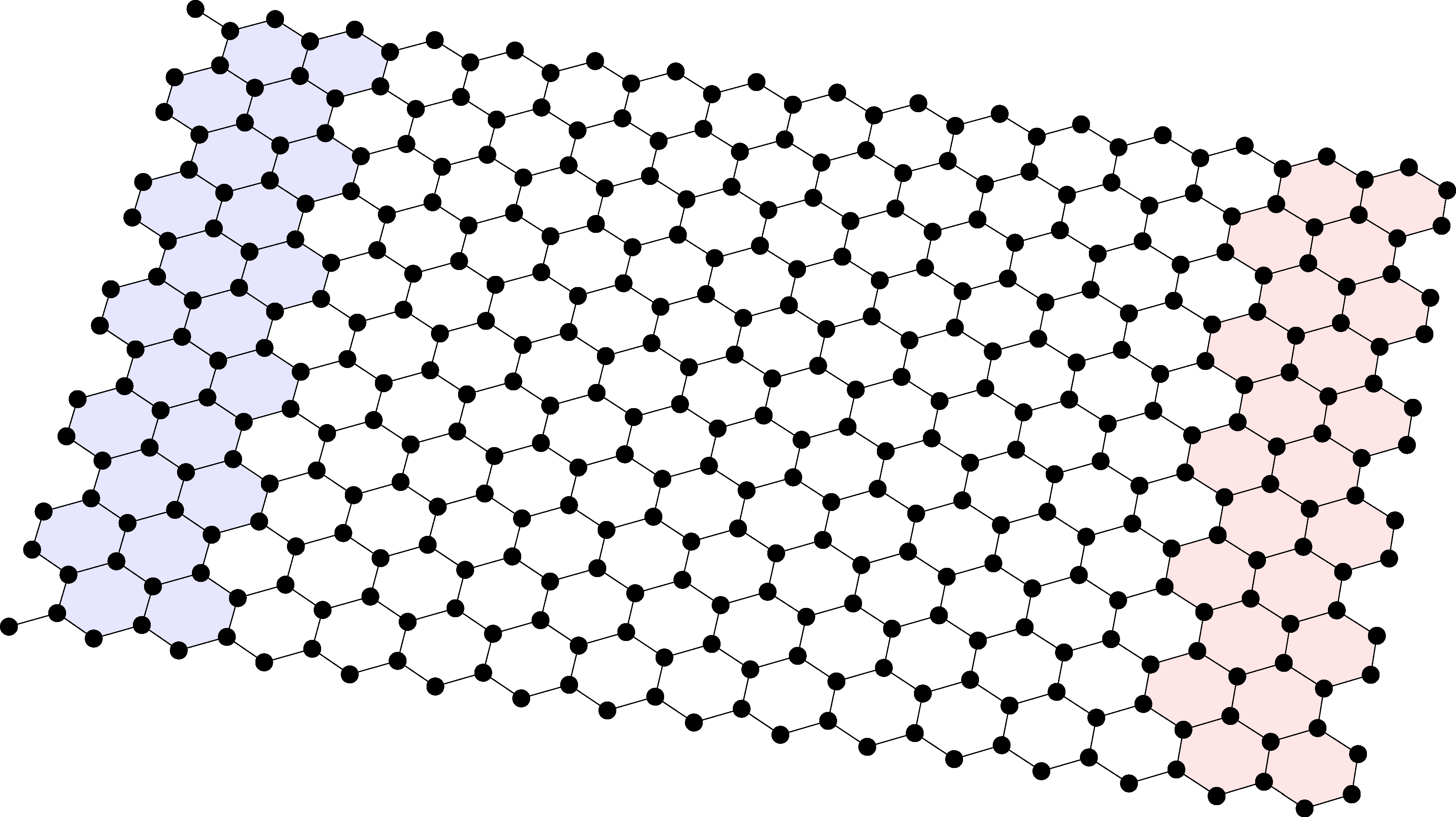}}\\
	\subfloat[Graphene nano-constriction]{\includegraphics[width=0.35\textwidth]{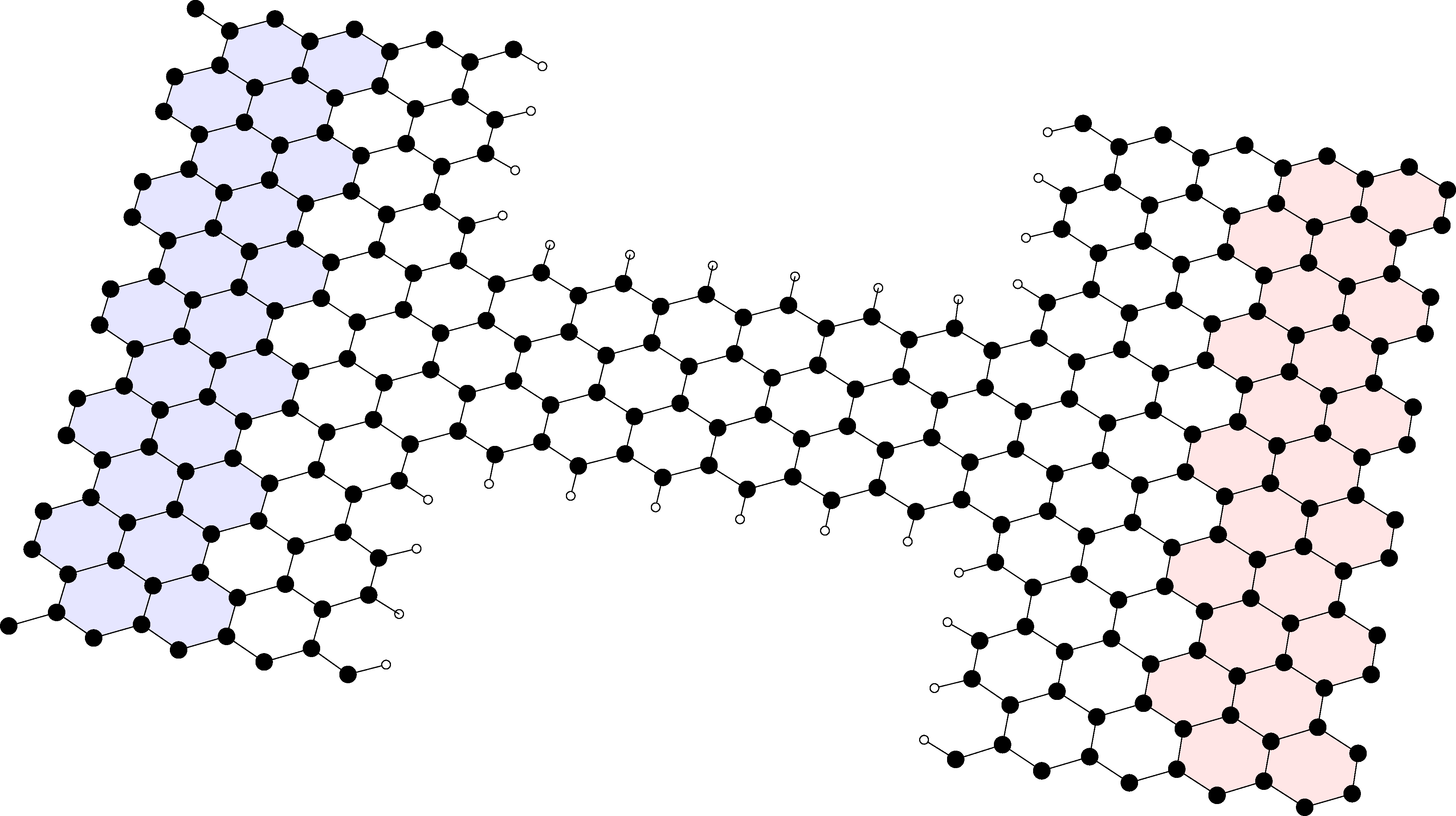}}\\
	\subfloat[Kinked graphene]{\includegraphics[width=0.35\textwidth]{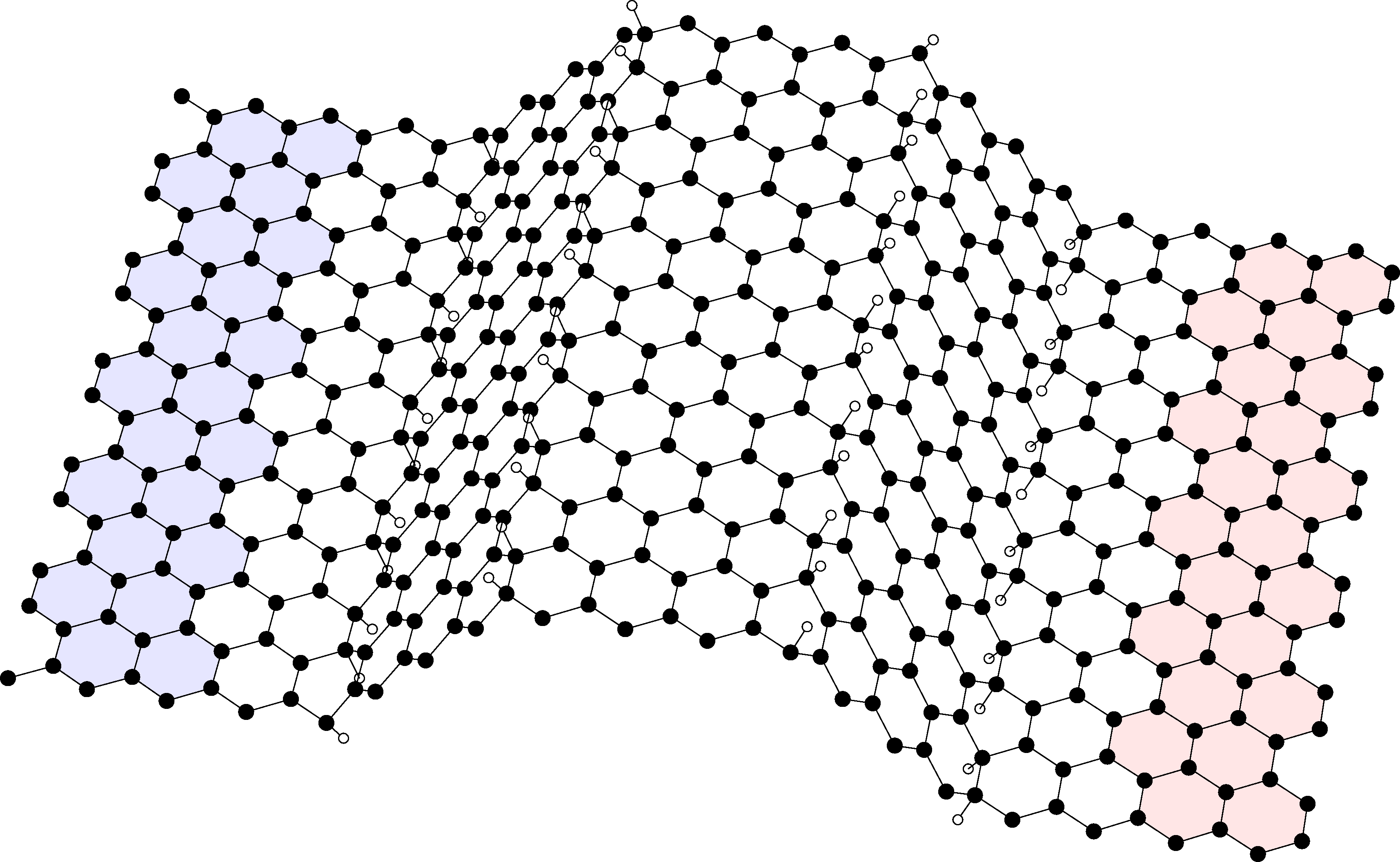}}
	\caption{(Colour online) Structures used as example cases: (a) a graphene sheet consisting entirely of carbon atoms, (b) a graphene hydrogen-passivated nano-constriction, and (c) a kinked graphene sheet decorated with hydrogen along the kink lines. The left and right electrodes have been highlighted with blue and red, respectively. Minimal unit-cells in the transverse transmission direction have been used due to the periodicity. Carbon atoms are shown in black, while hydrogen atoms are shown in white.}
	\label{fig:structures}
\end{figure}

The usefulness of the presented algorithm is showcased by considering transmission calculations through the simulated nano-systems shown in \figref{fig:structures}: (a) a pristine graphene sheet, (b) a graphene nano-constriction \cite{gunst_2013_phonon}, and (c) hydrogenated kinked graphene \cite{rasmussen_electronic_2013}. The shown structures have minimal unit-cells in the transverse transmission direction due to periodic boundary conditions, and carbon atoms are shown in black while hydrogen atoms are shown in white.

\begin{figure*}[t]
	\includegraphics{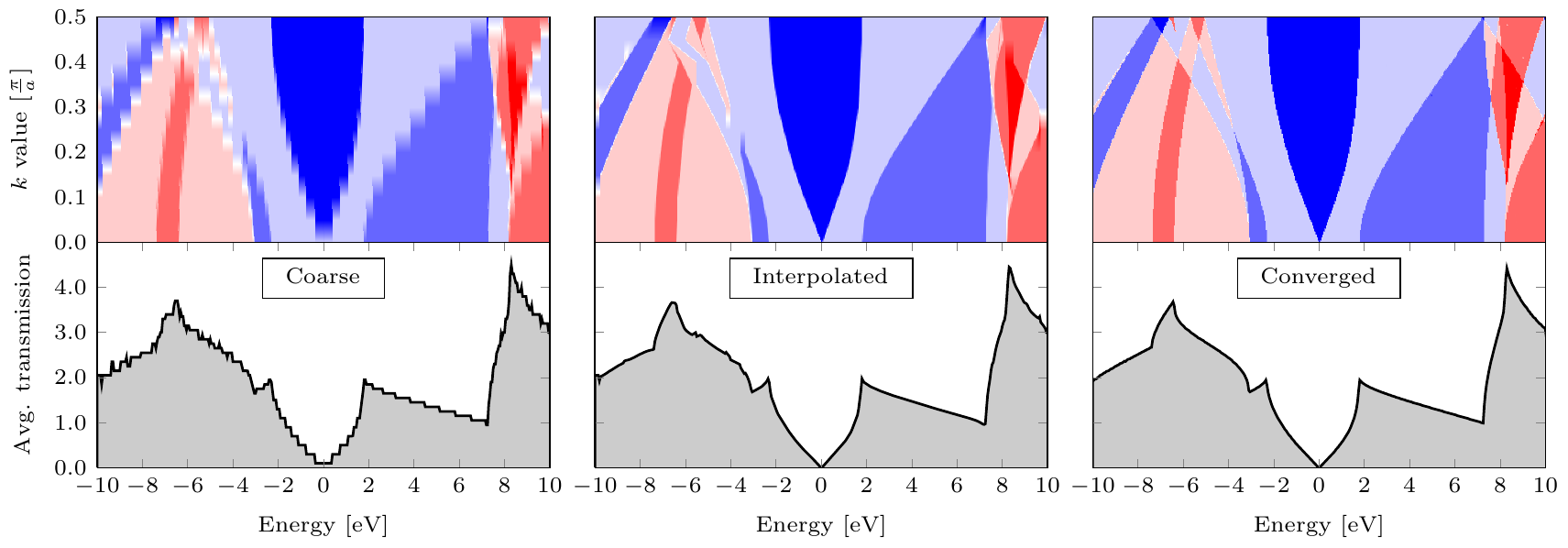}
	\includegraphics{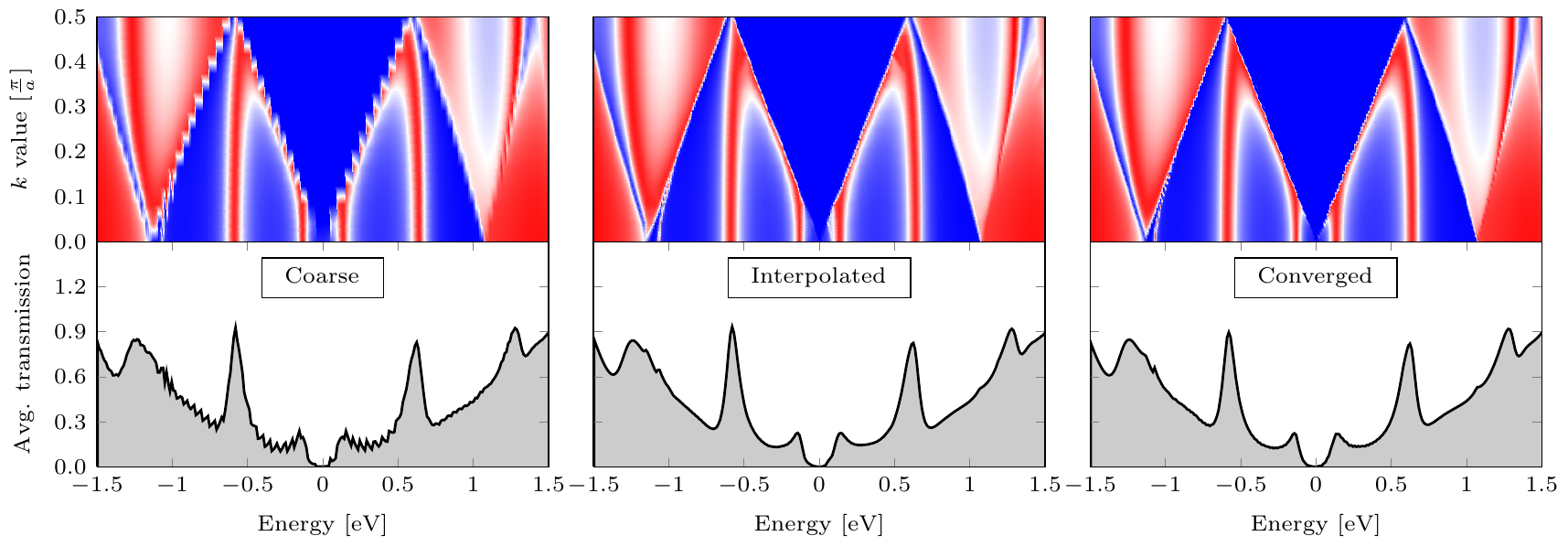}
	\includegraphics{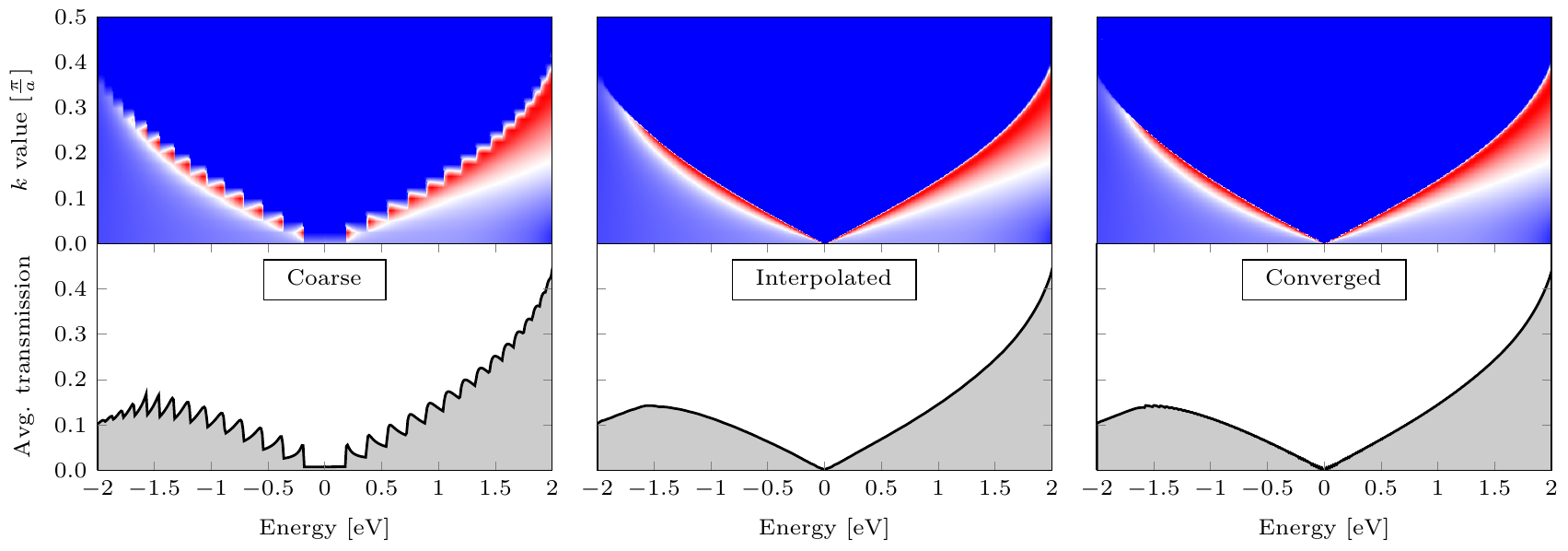}
	\caption{(Colour online) Algorithm applied to raw transmission data for the graphene-based systems in \figref{fig:structures}. The raw data (left column) is interpolated (middle column) and compared to the fully converged transmission curve (right column). The upper window of each subfigure is the full transmission for each $k$-value, while the lower window shows the averaged value.}
	\label{fig:transmission}
\end{figure*}

The transmission through the graphene sheet in \figref{fig:structures}a is calculated with the \textsc{TranSiesta} simulation suite, which utilizes a localized basis set. For the present work a DZP basis set is used in conjunction with a mesh cut-off of 300~Ry, a force tolerance of $0.02$~eV/\AA, as well as a Monkhorst-Pack grid of $24\times5$ ensuring absolute convergence. Exchange and correlation is described using the PBE GGA functional \cite{perdew_1996_generalized}. A minimal transverse unitcell is used due to periodic boundary conditions, while an energy window of $\pm 10$~eV around the Fermi energy is considered both for a crude transverse transmission $k$-grid ($N_k\!=\!20$) and a fully converged $k$-grid ($N_k\!=\!600$). The coarse transmission spectrum is interpolated using the presented interpolation scheme, as can be seen in the upper row in \figref{fig:transmission}. The upper window of each column describes the full transmission versus $k$-point. A good agreement is seen between the interpolated and the converged data sets. A few places the interpolation guesses incorrectly (for instance around $E=-4$~eV for $k=0.25\frac{\pi}{2}$). These occur since the input data is too coarse. However, despite the few differences between interpolated data and converged data, the averaged curves have a remarkable resemblance and the mean deviation is below 2\%, as seen in \figref{fig:introduction}b.

The transmission through the graphene nano-constriction and kinked graphene shown in Figs.~\ref{fig:structures}b and \ref{fig:structures}c have been extracted from the original datasets (see Refs.\cite{gunst_2013_phonon,rasmussen_electronic_2013}), and are shown in the middle and lower windows in \figref{fig:transmission}. As with the pristine graphene example we see a remarkable resemblance to the converged data set. The discrepancies are negligible, which stems from the fact that a finer $k$-point sampling has been performed in the original data. In the three cases we obtain speed-ups of approximately 5, 6, and 8, for the pristine graphene sheet, the graphene nano-constriction, and the kinked graphene sheet, respectively. Thus, we have demonstrated that by applying a simple post-processing interpolation scheme we can speed up convergence of roughly an order of magnitude.

\phantom{ }

\section{Comments and conclusion}
\label{sec:conclusion}
We have presented a simple post-processing interpolation scheme which speeds up transmission calculations on nano-structured materials. The algorithm uses a shortest-path solver to determine the optimal interpolation of a set of $k$-dependent transmission curves, which ultimately can be summed to obtain a smoothed average transmission.

We note that as a post-processing tool the algorithm relies on the quality of the original data. Since this data is used as a base for the interpolation any fluctuations will be present in the interpolated data and thus propagate to the final result. The present implementation of the shortest-path solver is based on Dijkstra's algorithm which is stable but very slow\cite{dijkstra_1959_anote}. A far quicker implementation would be to use a heuristic to guide the shortest-path search in the distance landscape, thus changing the solver to an industry standard algorithm known as an A$^*$-search\cite{hart_1968_formal}. However, due to the complexity of the input data the construction of such a heuristic is not a straight-forward task.

By considering three sample cases we have demonstrated that it can speed up calculations by roughly an order of magnitude. We have illustrated the method using electron transport through graphene nano-structures and $k$-point averaging of transmission functions. However, the method is generally applicable also to phonon transport and to other functions such as density of states or other types of interpolation parameters such as electrostatic gating etc.

Our interpolation scheme can easily be implemented in already existing code. We provide a sample \textsc{MatLAB} code online that can read and interpolate data obtained from \textsc{TranSiesta} and \textsc{TBTrans}\cite{brandbyge_density_2002}, which are built on-top of the \textit{ab-initio} software package \textsc{Siesta}\cite{soler_siesta_2002}.

\begin{acknowledgments}
The authors thank the Danish e-Infrastructure Cooperation (DeIC) for providing computer resources, and the Lundbeck foundation for support (R95-A10510).
The Center for Nanostructured Graphene is sponsored by the Danish National Research Foundation.
\end{acknowledgments}

\nocite{*}
\bibliographystyle{plain}
\bibliography{references}

\end{document}